\documentclass[fleqn,usenatbib]{mnras}

\usepackage{newtxtext,newtxmath}

\usepackage[T1]{fontenc}

\DeclareRobustCommand{\VAN}[3]{#2}
\let\VANthebibliography\thebibliography
\def\thebibliography{\DeclareRobustCommand{\VAN}[3]{##3}\VANthebibliography}

\usepackage{graphicx}	
\usepackage{amsmath}	
\usepackage[caption=false]{subfig}

\usepackage[svgnames]{xcolor}

\title[Angular Momentum Loss Rates in Be Stars Determined by the Viscous Decretion Disc Model] {Angular Momentum Loss Rates in Be Stars Determined by the Viscous Decretion Disc Model}

\defcitealias{granada2013}{G13}
\defcitealias{ghoreyshi2018}{G18}
\defcitealias{rimulo2018}{R18}

\author[M. R. Ghoreyshi et al.]{
M. R. Ghoreyshi,$^{1}$\thanks{E-mail: sghoreys@uwo.ca}
C. E. Jones,$^{1}$
and A. Granada$^{2, 3}$
\\
$^{1}$Department of Physics and Astronomy, The University of Western Ontario, London, ON N6A 3K7, Canada\\
$^{2}$Universidad Nacional de Rio Negro, Sede Andina, CITECCA, Anasagasti 1461, S.C. de Bariloche, Argentina\\
$^{3}$Consejo Nacional de Investigaciones Cient\'{i}ficas y T\'{e}cnicas, Argentina
}

\date{Accepted XXX. Received YYY; in original form ZZZ}

\pubyear{2022}

\begin{document}
\label{firstpage}
\pagerange{\pageref{firstpage}--\pageref{lastpage}}
\maketitle

\begin{abstract}

Circumstellar discs around Be stars are formed by the material ejected by the central star. This process removes excess angular momentum from the star as viscosity facilitates the mass and angular momentum transfer within the disc and its growth. The angular momentum loss rates (AMLR) of Be stars is a subject of debate in the literature. Through the modelling of the disc formation and dissipation phases observed from Be stars, their average AMLR can be determined and this is the goal of this work. We use the viscous decretion disc (VDD) model to provide a range of the average AMLR for Be stars and compare these rates with predicted values from the literature. We explore the reasons for discrepancies between the predicted values of average AMLR using the VDD and Geneva stellar evolution (GSE) models that were previously reported in literature and find that the largest differences occur when Be stars are rotating below their critical speeds. We show that the time over which the mass reservoir builds up is inversely proportional to the average AMLR. Also, we determine a revised value of the average AMLR for the Galactic Be star $\omega$ CMa of $4.7\times 10^{36}\,\mathrm{g\,cm^2\,s^{-2}}$, which is in better agreement with the values expected for a typical B2 type star. Finally, the effect of disc truncation due to the presence of a companion star is investigated and we find that this has a minimal effect on the average AMLR.

\end{abstract}

\begin{keywords}
stars: emission-line, Be -- individual: $\omega$ CMa -- massive -- mass-loss -- rotation
\end{keywords}



\section{Introduction}
\label{sect:intro}

Members of a subclass of main-sequence B-type stars are called Be stars because their spectrum occasionally show one or more hydrogen emission lines \citep{jaschek1981, collins1987}, specifically, the first members of the Balmer series \citep{rivinius2013a}. These emission lines originate in an equatorial, dust-free, near-Keplerian disc. Similar to normal B-type stars, the initial mass of Be stars varies from $\approx$\, 3.2\,M$_{\odot}$ to $\approx$\,17.5\,M$_{\odot}$ \citep{faes2015}. However, their rotational speed is significantly greater, from about half \citep[usually for early-types;][]{cranmer2005} to 95\% \citep[usually for late-types;][]{domiciano2012} of their critical velocity. Due to this fast rotation, the effective gravity near the equator of the Be stars is weaker compared to their slowly rotating counterparts. The fast rotation combined with other mechanism(s), perhaps nonradial pulsations \citep[NRP; e.g.,][]{rivinius2003, kee2016b}, helps to launch equatorial material into orbit. Most of the ejected gas particles fall back on the stellar surface but some gain enough angular momentum (AM) to remain in orbit and form a disc \citep{okazaki2002}. Viscous torque facilitates the transfer of mass within the disc out to larger radii, carrying AM away from the star \citep{carciofi2009, carciofi2012} as the disc grows in size. This process forms the foundation of the viscous decretion disc (VDD) model \citep{lee1991} that has been successfully used in modelling a variety of Be stars \citep[e.g.,][]{carciofi2009, carciofi2010, carciofi2012, klement2015, klement2017, klement2019, faes2016, baade2018a, rimulo2018, ghoreyshi2018, ghoreyshi2021, dealmeida2020, suffak2020, marr2021}.

Fast rotation not only causes lower effective gravity in the equatorial region and an oblate stellar shape, but it is an important factor in stellar evolution. Surface angular rotation enhancement is the consequence of core contraction and internal AM redistribution during the evolution of a Be star \citep[][hereafter G13]{ekstrom2008, granada2013}. Be star discs provide a natural mechanism for removing a significant amount of AM excess from the outer layers of the star \citep{krticka2011} while it evolves through the main sequence. \citetalias{granada2013} used the Geneva stellar evolution (GSE) code to estimate the theoretical AM loss rate (AMLR) for Be stars during main sequence evolution. They assumed that a steady-state viscous decretion disc forms whenever the outer equatorial layers of the star reaches $W \geq 0.88$ (or equivalently $\omega \geq 0.99$), where $W=v_\mathrm{rot}/v_\mathrm{orb}$ and $\omega = \Omega_\mathrm{rot}/\Omega_\mathrm{orb}$ are the linear and angular velocity ratios, respectively. $v_\mathrm{rot}$ and $\Omega_\mathrm{rot}$ are the linear and angular rotational speeds. $v_\mathrm{orb}$ and $\Omega_\mathrm{orb}$ are the linear and angular orbital speeds of the particles. By modelling the photometric data of a group of Be stars in the small Magellanic cloud (SMC) using the VDD model, \citet[][hereafter R18]{rimulo2018} found that their predictions of AMLR are much smaller than the values suggested by the GSE model. Later, \citet[][hereafter G18]{ghoreyshi2018} calculated the average AMLR of the Galactic Be star $\omega$ CMa by modelling its 34-year photometric $V$-band data and confirmed that the AMLR are one order of magnitude smaller than the predictions from the GSE model.

The main goal of this work is to answer two questions: 1) what is (are) the reason(s) for the difference in predictions of the GSE and VDD models and 2) what is the potential range of average AMLR for Be stars as predicted by the VDD models? To address these questions, we need to understand the details of the models and the assumptions that were adopted. It is important, even for a broadly tested model such as the VDD model, to ensure that all assumptions are well understood and applied.

To determine the AMLR for Be stars, it is necessary to consider the full range of sub-spectral types. The studies of \citetalias{ghoreyshi2018} and \citetalias{rimulo2018} included early-type Be stars with a minimum mass of 6.3\, M$_{\odot}$. Here, we aim to broaden the results reported in the literature by applying the VDD models to a full range of Be star sub-spectral types. First, our study uses a restricted range of input parameters and then we randomly select combinations of input parameters for a large number of stars from 3.2 to 17.5M$_{\odot}$. 

Section~\ref{sect:models} provides details about the GSE and VDD models. Section~\ref{sect:controlled} presents the results of the average AMLR determination by using the VDD model for the Be stars with controlled combinations of stellar and disc parameters and compares to the results of \citetalias{granada2013}. In Section~\ref{sect:statistical}, we repeat our analysis with randomly selected combinations of stellar and disc parameters for a large number of sample Be stars. Section~\ref{sect:discussion} discusses our results. Finally, in Section~\ref{sect:conclusions} our conclusions and plans for future work are presented.


\section{Theoretical Models}
\label{sect:models}

 In this Section, the major properties of the VDD and GSE models are outlined, including the main methods and approximations used to obtain the average AMLR.

\subsection{Viscous Decretion Disc Model}
\label{subsect:vdd}

Foundations for the VDD model were developed by \cite{lee1991} and then the model was further advanced by \cite{porter1999, bjorkman2001, okazaki2001, bjorkman2005, krticka2011}. In Be stars, the effective gravity at the polar regions is higher than the equatorial regions because of the oblate stellar shape, as explained above. This latitudinal dependency of effective gravity creates a corresponding latitudinal dependence on the flux, so that the poles are brighter (i.e., hotter) and the equator is darker (i.e., cooler), according to the \cite{vonzeipel1924} theorem. This can be expressed as follows in its original formulation, which is applicable to a purely radiative envelope:
\begin{equation}
\centering
\label{eq:vonzeipel}
    T_\mathrm{eff}(\theta) \propto g^{\beta}_\mathrm{eff}(\theta),
\end{equation}
with $\beta=0.25$. Later, \citet{espinosa2011} demonstrated that the value of $\beta$ is a function of the stellar rotational rate; from 0.21 for $W=0.50$ to 0.15 for $W=0.95$.

Be star discs alter the stellar light that is observed. Variations in the system's observables are caused by changes in the geometrical shape, density, and orientation of the disc to the line of sight. The disc is fed by gas expelled from the star according to the VDD model. As a result of the viscous diffusion of material, the matter injected into the disc can spread both inwards and outwards. 

The mass flux, $\dot{M}_\mathrm{disc} (r, t)$, and the AM flux, $\dot{J}_\mathrm{disc} (r, t)$ are the relevant disc quantities in this study. These characterise the flow of matter and AM in the disc as a function of inner and outer disc boundary conditions; e.g., a stellar varying mass loss rate, and a disc that is truncated by a binary companion, respectively. A positive sign for these values means the flow is outward (i.e., when the star loses mass), and a negative sign corresponds to an inward flow (i.e., when the material falls back onto the star). The azimuthal velocity, $v_{\phi}$, of the gas with subsonic outflow is Keplerian, i.e., $v_{\phi} \propto r^{-0.5}$ \citep[e.g.,][]{pringle1981}, according to a common assumption in VDDs for Be stars. Under this assumption, the mass flux is related to the stationary disc surface density, $\Sigma(r, t)$, by the mass conservation relation given by \cite{balbus2003} and \citetalias{rimulo2018} as
\begin{equation}
    \label{eq:mdot_disc}
    \dot{M}_\mathrm{disc}(r,t) = 2\pi r \Sigma(r, t) v_r = -4\pi \sqrt{\frac{r}{GM_*}} \frac{\partial}{\partial r}\left(\alpha c_\mathrm{s}^2 r^2\Sigma(r, t)\right),
\end{equation}
where $v_r$ is the radial speed, $G$ is the universal gravitational constant, $\alpha$ is the viscosity parameter of \cite{shakura1973}, $c_\mathrm{s}$ is the isothermal sound speed, given by $c_\mathrm{s}^2=kT_\mathrm{disc}/\mu m_\mathrm{H}$, and $r$ is the distance from the center of the star which is located a the origin of a cylindrical coordinate system. At regions relatively close to the stellar photosphere, the temperature profile of the disc, $T_{\rm disc}$, is well represented by a function associated with a black body reprocessing disc \citep{adams1987}. The temperature profile follows this function as long as the disc is optically thick, after which it increases to a constant temperature of about 60\% of stellar effective temperature, $T_\mathrm{eff}$ \citep{carciofi2006a}. This approximation for the $T_\mathrm{disc}$ is used throughout the paper.

\cite{shakura1973} introduced the Shakura—Sunyaev viscosity parameter, $\alpha$, as a dimensionless parameter to explain the kinematic viscosity, and assumed the viscosity, $\nu$, to be given by $\nu=v_\mathrm{tur}l=\alpha c_\mathrm{s} H$. This is the viscosity prescription that is commonly used in the literature and is adopted here. The function of $\alpha$ is to link the turbulence scale to the vertical scale of the disc, $H$.
The turbulence is composed of eddies (vortices) and $l$ is the size scale of the largest eddies, while $v_\mathrm{tur}$ is the ``turnover'' velocity of the eddies. Because the greatest eddies can only be about the size of the disc scale height, we set $l = H$. It can be assumed that the outflow velocity is of the order or smaller than the sound speed ($c_\mathrm{s}$). Otherwise, the turbulence would be supersonic, and the eddies would disintegrate into a succession of shocks. Therefore, $\alpha$ is restricted to the values equal to 1.0 or smaller.

Substituting $\dot{M} =\dot{J}/\sqrt{GM_*r}$ into equation 27 of \cite{bjorkman2005}, the conservation of AM flux for a stationary disc, can be written as
\begin{equation}
\label{eq:jdot_disc}
\dot{J}_\mathrm{disc}(r,t) = 2\pi r \Sigma(r, t) v_r \sqrt{GM_*r} + 2\pi \alpha c_\mathrm{s}^2 r^2\Sigma(r, t)\,,
\end{equation}
where the first term is the AM flux carried by the gas's radial motion, and the second term is the AM flux due to the torque created by the viscous force. 

The following diffusion-like equation describes the time evolution of the surface density \citep{papaloizou1995}, which holds in the thin disc approximation ($c_\mathrm{s}^2\ll GM_*/r$):

\begin{equation}
\label{eq:sigmadot}
\frac{\partial{\Sigma}}{\partial{t}}=\frac{2}{r} \frac{\partial}{\partial{r}} \left\{\sqrt{r}\frac{\partial{}}{\partial{r}}\left[\frac{\alpha c_\mathrm{s}^2}{\sqrt{GM_*}} r^2 \Sigma(r, t)\right]\right\}+S_\Sigma,
\end{equation}
where $S_\Sigma$ is the rate of mass injection from the star into the disc per unit area.

Following \citetalias{ghoreyshi2018} and \citetalias{rimulo2018}, it was assumed that the star injects mass in a Keplerian orbit at a given rate, $\dot{M}_\mathrm{inj}$, at a radius very close to the surface of the star, $R_\mathrm{inj}$, in the equatorial plane (see below). Because there is observable evidence pointing to asymmetric mass loss \citep[e.g., the short-term V/R variations of $\eta$ Cen,][]{rivinius1997} or matter being ejected at higher latitudes \citep[e.g.,][]{stefl2003a, ghoreyshi2021}, this assumption may be overly simplistic for some Be stars. However, this is not a concern in this study because, due to viscous diffusion and orbital phase mixing, the gas loses memory of the injection procedure after a few orbital cycles. Hence, $S_\Sigma$ is assumed to be:
\begin{equation}
\label{eq:s_sigma}
S_\Sigma=\frac{\dot{M}_\mathrm{inj}}{2\pi}\frac{\delta(r-R_\mathrm{inj})}{r},
\end{equation}
as given by \citetalias{rimulo2018}.

Finally, we assume torque-free boundary conditions at the stellar equator, $R_\mathrm{eq}$, and at large outer disc radii, $R_\mathrm{out}$. As the second term in Equation~\ref{eq:jdot_disc} shows, this is accomplished by setting $\Sigma=0$ at these boundaries. The outer boundary, for instance, may indicate the limiting radius of the disc as a result of a binary companion (e.g., \citealt{okazaki2002}) or photo-evaporation of the disc (e.g., \citealt{kee2016a}). Note that $R_\mathrm{out}$ is not a physical outer edge of the disc. Rather, it simply satisfies the torque-free condition at the outer boundary. Later, we discuss how this parameter may affect our models.

The conservation of AM through the system and the outside medium, i.e., the interstellar medium, can be written by
\begin{equation}
\label{eq:jdot_star}
\dot{J}_*(t)+\frac{\mathrm{d}\phantom{t}}{\mathrm{d}t}\int_{R_\mathrm{eq}}^{R_\mathrm{out}}\sqrt{GM_*r}\Sigma(r,t) 2\pi r\mathrm{d}r+\dot{J}_\mathrm{disc}(R_\mathrm{out},t)=0\,,
\end{equation}
as follows \citetalias{rimulo2018}, where the variation rates of the AM in the star, the disc, and the outer medium are represented by the first, second, and third terms, respectively. Solving Equation~\ref{eq:sigmadot} for $\Sigma(r,t)$ yields the second and third terms of Equation~\ref{eq:jdot_star}, allowing us to calculate the AM lost by the star.

In a steady-state mass feeding scenario\footnote{It is important to distinguish between ``steady-state disc'' and ``steady mass or AM injection rate''. The former refers to a disc whose properties do not vary with time. The second term refers to a constant mass and AM injection rate.}, AM is injected at a constant rate $\sqrt{GM_* R_\mathrm{inj}}\dot{M}_\mathrm{inj}$ at the radius $R_\mathrm{inj}$. This AM is divided into a constant AM flux inwards, for $r<R_\mathrm{inj}$, and a constant AM flux outwards, for $r>R_\mathrm{inj}$. Since in a steady-state disc, the AM (second term of Equation~\ref{eq:jdot_star}) is constant in time, therefore the outward flux in the region $r>R_\mathrm{inj}$ is equal to the AMLR of the star. By substituting equation 11 of \citetalias{rimulo2018} into their equation 13, we derive the following equation for the AMLR of the star:
\begin{equation}
\label{eq:jdot_steady}
-\dot{J}_{*,\mathrm{std}}=\Lambda \sqrt{GM_*R_\mathrm{eq}}\dot{M}_\mathrm{inj}\left(\sqrt{\frac{R_\mathrm{inj}}{R_\mathrm{eq}}}-1\right)\,,
\end{equation}
where $\Lambda=1/(1-\sqrt{R_\mathrm{eq}/R_\mathrm{out}})$ is a number usually just slightly larger than 1, as $R_\mathrm{out}\gg R_\mathrm{eq}$. 
The steady-state AMLR, therefore, depends very little on the outer radius of the disc (through the factor $\Lambda$), whose value is approximately known for a few Be stars, and completely unknown for most \citep{klement2017}. In the simulations presented in Section~\ref{sect:controlled}, we set $R_\mathrm{out}=1000R_\mathrm{eq}$, while in Section~\ref{sect:statistical}, $R_\mathrm{out}$ is a model input parameter that changes randomly.

Following \cite{bjorkman2005} and \citetalias{rimulo2018}, the steady-state surface density for radii larger than $R_\mathrm{inj}$ is given by
\begin{equation}
\label{eq:sigma_steady}
\Sigma_\mathrm{std}(r)=\frac{-\dot{J}_{*,\mathrm{std}}}{2\pi\alpha c_\mathrm{s}^2}\frac{1}{r^2}\left(1-\sqrt{\frac{r}{R_\mathrm{out}}}\right)\,,\,R_\mathrm{inj}\leq r \leq R_\mathrm{out}\,,
\end{equation}
from which we see that the disc density, a physical quantity that can be estimated, for example, by modelling the spectral energy distribution profile, scales with $\dot{J}_{*,\mathrm{std}}/\alpha$, or, alternatively, with $\dot{M}_{\rm inj}$/$\alpha$. This highlights a key feature of VDDs: the injection rates can only be determined if $\alpha$ is known. Note that $\Sigma_\mathrm{std}$ and $\dot{J}_{*,\mathrm{std}}$ are just another way of expressing the rate of mass and AM injection into the disc and their index ``std'' is to emphasize that these parameters are defined in the steady-state limit. Hereafter, the mass and AM injection rates will be expressed in terms of $\Sigma_0$, $\dot{M}_{\rm inj}$, or $\dot{J}_{*,\mathrm{std}}$, interchangeably, as these three quantities are connected by Equations~\ref{eq:jdot_steady} and~\ref{eq:sigma_steady}.

\begin{figure}
    \centering
    \includegraphics[width=\linewidth]{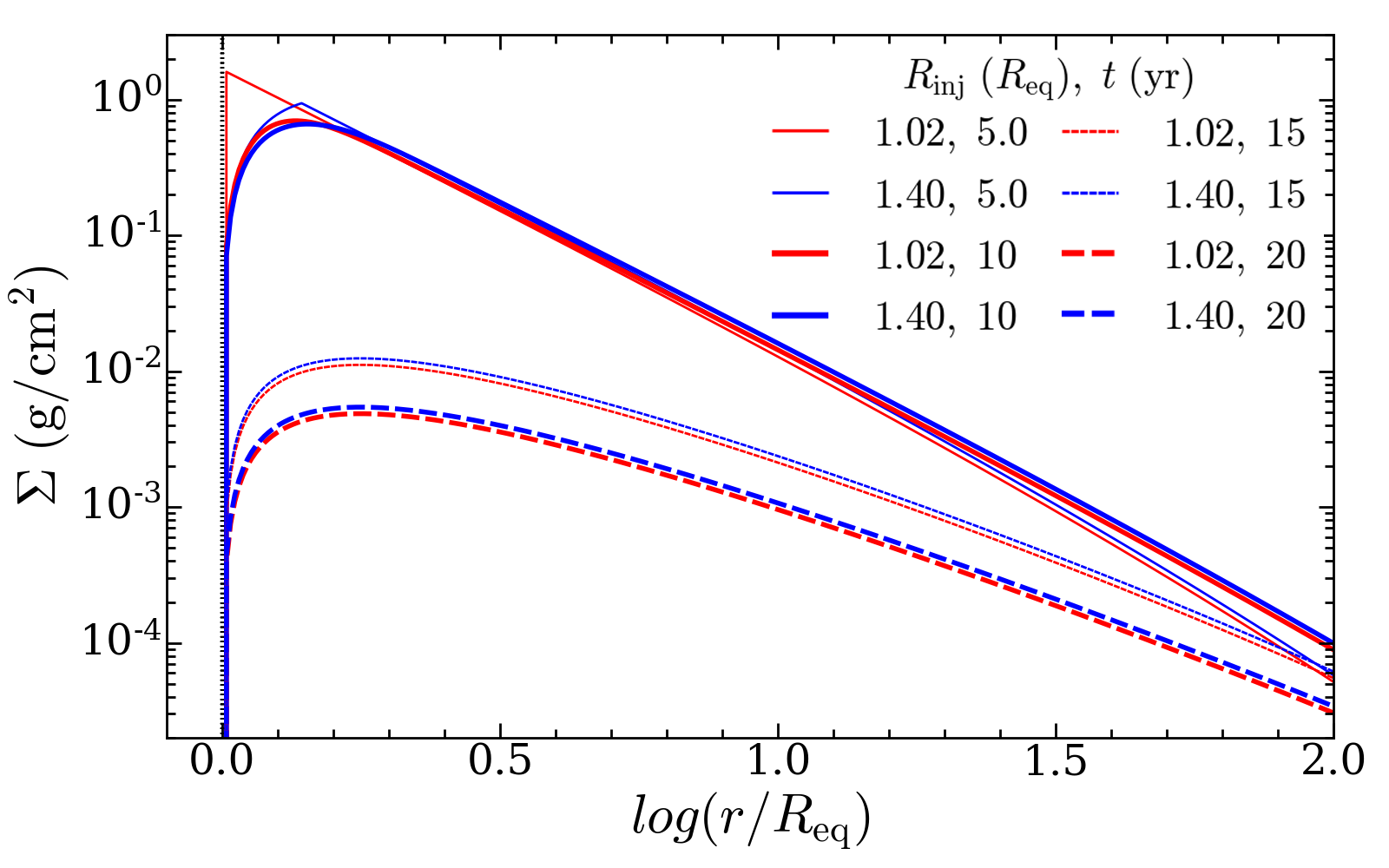}
    \caption{Effect of the mass injection point on the surface density ($\Sigma$) profile. The disc was built in 10 years, followed by a 10-year disc dissipation. $\Sigma$ is a function of distance from the star for two different injection points, as indicated, at four different epochs: (a) at the middle of disc formation phase ($t=5.0\,\mathrm{yr}$), (b) at the end of disc formation phase ($t=10.0\,\mathrm{yr}$), (c) at the middle of disc dissipation phase ($t=15.0\,\mathrm{yr}$), and (d) at the end of disc dissipation phase ($t=20.0\,\mathrm{yr}$). The red and blue lines show $\Sigma$ at $R_\mathrm{inj}=1.02 R_\mathrm{eq}$, and $R_\mathrm{inj}=1.40 R_\mathrm{eq}$, respectively.}
    \label{fig:injection_point}
\end{figure}

The time-variable source term of Equation~\ref{eq:s_sigma} drives the dynamical evolution of the disc surface density (Equation~\ref{eq:sigmadot}). It has been verified \citepalias{rimulo2018} that the solution of Equation~\ref{eq:sigmadot} for $r > R_\mathrm{inj}$ is negligibly affected by variations of both $\dot{M}_\mathrm{inj}(t)$ and $R_\mathrm{inj}/R_\mathrm{eq}$ as long as the product $\dot{M}_\mathrm{inj}(t)(\sqrt{R_\mathrm{inj}/R_\mathrm{eq}}-1)$ is kept unchanged. Figure~\ref{fig:injection_point} illustrates this point. It shows the solution of $\Sigma(r,t)$ at a certain time, $t$ (in different stages of the disc formation and dissipation) for the case of a disc that is being built-up for 10 years from a discless state and then dissipates for another 10 years. The two curves have different values of $\dot{M}_\mathrm{inj}$ and $R_\mathrm{inj}/R_\mathrm{eq}$ (illustrated by red, and blue lines) but the product $\dot{M}_\mathrm{inj}(\sqrt{R_\mathrm{inj}/R_\mathrm{eq}}-1)$ is the same for each. Figure~\ref{fig:injection_point} shows that $\Sigma(r,t)$ only differs significantly in the small interval $R_\mathrm{eq} < r < R_\mathrm{inj}$ which contributes negligibly to the total emission of the disc as long as $R_\mathrm{inj}/R_\mathrm{eq}$ is close to 1. This is a reasonable assumption since the effect is small and is only relevant when the disc is being formed. Since the radius $R_\mathrm{inj}$ is not well known, it follows that the mass injection into the disc is best described by the quantity defined in Equation~\ref{eq:jdot_steady}, which approximately represents the net AM injected into the disc. 

At outburst, $\dot{J}_{*,\mathrm{std}}$ is non-zero, which results in an outflowing ($v_{r}>0$, $\dot{M}_\mathrm{disc} > 0$) decretion disc. The disc is steadily formed from the inside out during this phase \citep{haubois2012}. Quiescence is typically defined as a state in which $\dot{J}_{*,\mathrm{std}}=0$. In this phase, the inner disc passively diffuses inward ($v_{r}<0$, $\dot{M}_\mathrm{disc} < 0$) while the outer disc remains outflowing \citep[see figure 10 of][]{haubois2012}. 


Equation~\ref{eq:jdot_star}, where $\dot{J}_{*}$ represents the rate of AM being lost by the star (AMLR), can also be written as follows:
\begin{equation}
-\dot{J}_{*}  = \dot{M}_\mathrm{inj}\sqrt{GMR_\mathrm{inj}} + \dot{M}_\mathrm{eq}\sqrt{GMR_\mathrm{eq}} \,,
\label{eq:jdot_star2} 
\end{equation}
which means that AM is injected in a ring with an average radius, $R_\mathrm{inj}$, at a rate given by the first term of the equation. A large fraction of the injected AM is re-accreted onto the stellar surface at the equatorial region at a rate given by the second term of the equation. The remaining fraction of the injected AM is the instantaneous AMLR from the star into the disc, $\dot{J}_{*}$. Simulations also show that $\dot{J}_{*}$ is close to the steady-state AMLR, $\dot{J}_\mathrm{*,std}$, given by Equation~\ref{eq:jdot_steady}, at most times. The physical interpretation of Equation~\ref{eq:jdot_star2} is as follows: in order that some of the material injected at $R_{\rm inj}$ gain sufficient AM to be driven outwards to larger orbits, the material must receive the equivalent AM that falls back to the star.

In describing the injection from the star into the disc, we avoid using the two parameters $\dot{M}_\mathrm{inj}$ and $R_\mathrm{inj}/R_\mathrm{eq}$, individually, which cannot be determined observationally. Instead, we use a function proportional to the product $\dot{M}_\mathrm{inj}(\sqrt{R_\mathrm{inj}/R_\mathrm{eq}}-1)$, which will generate a certain uniquely defined $\Sigma(r,t)$ profile that can be determined observationally. The realization that we should consider a function proportional to $\dot{M}_\mathrm{inj}(\sqrt{R_\mathrm{inj}/R_\mathrm{eq}}-1)$ comes from observing the form of the steady-state solution of Equation~\ref{eq:sigmadot}. To find the steady-state solution, we set $\partial{\Sigma}/\partial{t} = 0$ in Equation~\ref{eq:sigmadot}. This results in an ordinary differential equation for $\Sigma_\mathrm{std}(r)$ that is easy to solve analytically. In a steady-state AMLR, a fraction of the injected mass, $\Upsilon$, flows back to the stellar equatorial region, while the remaining fraction, $1-\Upsilon$, flows towards the outer boundary and exits the system. This fraction is given by \citetalias{rimulo2018}
\begin{equation}
\Upsilon =\frac{\sqrt{R_\mathrm{out}}-\sqrt{R_\mathrm{inj}}}{\sqrt{R_\mathrm{out}}-\sqrt{R_\mathrm{eq}}}\,,
\label{eq:upsilon} 
\end{equation}
from which we see that $\Upsilon$ is a number smaller than one, since $R_\mathrm{inj}/R_\mathrm{eq}$ is very close to one. Therefore, almost all the injected mass must return to the star. In addition, in a steady-state AMLR, the rate of mass re-accretion at the stellar equator is $\dot{M}_\mathrm{eq} = -\Upsilon\dot{M}_\mathrm{inj}$, and the rate of mass ejection from the disc through the outer boundary is $\dot{M}_\mathrm{out} = -(1-\Upsilon)\dot{M}_\mathrm{inj}$. We will see below that the steady-state AMLR has a very small dependence on this parameter, $\Upsilon$. The mass loss rate's dependence on $\Upsilon$, on the other hand, is not small, which justifies our choice of describing the mass injection by $\dot{J}_\mathrm{*,std}$. 

Substitution of Equation~\ref{eq:jdot_steady} into Equation~\ref{eq:sigma_steady} gives a new equation for the steady-state surface density that shows, although we may vary $\dot{M}_\mathrm{inj}$ and $R_\mathrm{inj}/R_\mathrm{eq}$, the function $\Sigma_\mathrm{std}(r)$ does not change as long as $\dot{J}_\mathrm{*,std}$ is kept constant. Equation~\ref{eq:jdot_steady} shows that the steady-state AMLR has a weak dependence on the outer radius $R_\mathrm{out}$, through the number $\Lambda$. For comparison, we provide an equation for the steady-state mass loss rate from the star,  $\dot{M}_\mathrm{*,std}$. It is given by $-\dot{M}_\mathrm{*,std} = -\dot{M}_\mathrm{out} = (1-\Upsilon)\dot{M}_\mathrm{inj}$, or
\begin{equation}
-\dot{M}_\mathrm{*,std} = \Lambda\dot{M}_\mathrm{inj}\sqrt{\frac{ R_\mathrm{eq}}{R_\mathrm{out}}}\left(\sqrt{\frac{R_\mathrm{inj}}{R_\mathrm{eq}}}-1\right)\,,
\label{eq:mdot_std} 
\end{equation}
from which we see that its dependency on $R_\mathrm{out}$ is significantly greater than that of $\dot{J}_\mathrm{*,std}$ on $R_\mathrm{out}$.

According to the pseudo-photosphere model, the size of optically thick part of Be star discs which contributes two thirds of their total emission flux, is largest at the longest wavelengths \citep[figure 10 in][]{vieira2015}. The formation loci of continuum emission for radio bands is approximately $30 R_\mathrm{eq}$ \citep[varies by disc density, figure 2 in][]{rivinius2013a}, while a good approximation for the outer radius of a non-truncated disc is given by the formula:
\begin{equation}
R_\mathrm{out} = 0.3(v_\mathrm{orb}/c_\mathrm{s})^2 R_\mathrm{eq}\,,
\label{eq:r_out} 
\end{equation}
\citep{krticka2011}. For typical Be stars, this formula suggests that $R_\mathrm{out}$ is between 200 and 600$R_\mathrm{eq}$. The spectral energy distribution of Be stars shows that their flux drops dramatically at these large distances. Consequently, the disc is too faint, and therefore, the outer radius $R_\mathrm{out}$, especially for non-truncated discs, cannot be determined observationally. In our dynamical models (in Section~\ref{sect:controlled}), $\Lambda = 1.03$, which is very close to one. If the true outer radius of the Be star system were $R_\mathrm{out} = 100 R_\mathrm{eq}$, meaning that $\Lambda = 1.11$, we would still have a negligible difference between our steady-state AMLR estimates and the actual AMLR. The steady-state mass loss rate estimations, on the other hand, would differ by more than a factor of three. Even if our Be star system was a member of a close binary system and the disc was truncated at $R_\mathrm{out} = 8R_\mathrm{eq}$, then we would have $\Lambda = 1.55$, meaning that we have a good determination of the steady-state AMLR, at least within an order of magnitude. But, in this case, the steady-state mass loss rate would differ by more than a factor of ten.


Finally, we note that $\dot{J}_{*,\mathrm{std}}$ changes over time \citepalias[see figure 11 in ][]{ghoreyshi2018}. Therefore, we define an average AMLR $\Tilde{\dot{J}}_{*,\mathrm{std}}$:
\begin{equation}
\label{eq:jdot_ave_integral}
\Tilde{\dot{J}}_{*,\mathrm{std}} = \frac{\int \dot{J}_{*,\mathrm{std}}dt}{\tau}\,,
\end{equation}
where $\tau$ is the total length of the time span over which the average AMLR is desired. For scenarios where $\dot{J}_{*,\mathrm{std}}$ has either zero (i.e., during disc dissipation phases) or a constant non-zero value (i.e., during disc formation phases, as in Sections~\ref{sect:controlled} and \ref{sect:statistical}). Equation~\ref{eq:jdot_ave_integral} can be simplified as:
\begin{equation}
\label{eq:jdot_ave}
\Tilde{\dot{J}}_{*,\mathrm{std}} = \mathrm{DC}\times\dot{J}_{*,\mathrm{std}}\,,
\end{equation}
where $\mathrm{DC}$ is the duty cycle, the time of star's active phase (when the disc builds up) divided by the total time of the active phase and quiescence phase (when the disc dissipates). Also, since we intend to compare the results of our calculations with the average AMLR of the Be star, $\omega$ CMa \citepalias{ghoreyshi2018}, Equation~\ref{eq:jdot_ave} is used throughout the paper to calculate a mean value for the AMLR with Equation~\ref{eq:jdot_steady} providing the value for $\dot{J}_{*,\mathrm{std}}$.



 
\subsection{Geneva Stellar Evolution Model}
\label{subsect:gse}

In order to compare our new determinations of AMLR based on the VDD model, with those obtained through stellar evolution calculations, we recall here the AMLR prescriptions considered in the Geneva Stellar Evolution (GSE) model. In this model, both, the surface and critical angular velocity change with the evolution of a star, resulting in a change in the rotation rate, $\omega=\Omega_{\rm surf}/\Omega_{\rm crit}$ as a function of time. It is possible for a star to approach the critical velocity during its life, even if it starts the Zero Age Main Sequence (ZAMS) with $\omega_{\rm ini}<1$. When the critical velocity is reached, the effective gravity at the equator of the star tends to zero, so any mechanism adding an outwards force, could cause an enhancement of the mass loss in the equatorial region. This is referred to as “mechanical mass loss”, and outer layers are removed while maintaining the surface rotation at the critical limit or slightly below. 
\citet{georgy2013} presented a grid of evolutionary models to explore the effects of stellar rotation in the mass domain between 1.7 to 15\,M$_{\odot}$, which correspond to early-A, -B, and late-O spectral types. Different metallicities and initial angular velocity between $0 < \Omega_{\rm ini}/\Omega{\rm crit}<0.95$ were considered. In their paper, the authors provide a full description on the angular momentum treatment within the star leading to mechanical mass loss when the critical limit is approached. For completeness, since we refer to these evolutionary models throughout the present article, we briefly describe  them here.

To account for the mechanical mass loss in the Geneva Stellar Evolution Code the amount of angular momentum that the model loses through radiative stellar winds during a time step are computed using estimated surface quantities such as radius, mass-loss rate, etc. Then, whether this mass wind mass loss is sufficient to keep the stellar surface rotating below the 0.99 critical angular velocity, $\Omega_{\rm crit}$, at the end of the time step is determined. This is important because numerical difficulties arise when approaching the critical limit. As the wind removes angular momentum from the stellar envelope, a new angular velocity distribution inside the star is built up so that the total angular momentum of the star will be decreased by the exact amount lost through winds. We refer the reader to \citet{georgy2013} for a detailed description of this process.

Once the structure of the envelope is recomputed, the surface angular velocity is also re-estimated at the end of each time-step. If this new value exceeds the defined critical limit, an amount of angular momentum is removed via mechanical mass loss. This prescription assumes that this mass is lost forever and does not re-accrete into the star.

 \citetalias{granada2013} investigated a subset of their grid, with equatorial velocities reaching the critical limit to investigate links to Be stars. Since the initial conditions for the computed models used solid rotation, the readjustment of the internal rotation profile led to a sharp decrease in the surface rotational velocity immediately after the ZAMS, which resulted in the rotation of the stellar surface below the original rates at this stage. This situation also prevented the authors to compute models with a larger angular momentum content at the ZAMS. This issue was circumvented by \citet{granada2014}, who computed evolutionary tracks using the same mass and metallicity range as \citet{georgy2013}, and also using a non solid rotation profile from the pre-main sequence with a larger angular momentum content, allowing them to compute evolutionary tracks with critical rotation over the entire main sequence. 

From the time averaged AMLR obtained by \citet{georgy2013} when critical rotation was attained, \citetalias{granada2013} and \citet{granada2014} used the parameterized expressions from \citet{krticka2011} to characterise the steady Keplerian discs that form around rapidly rotating stars. They provided tables with stellar masses at the ZAMS, averaged AMLRs, mass loss rates and outer disc radii, among other parameters, for stellar models with different metallicities and initial rotational rates. These are the quantities that we use for comparison of our modelling in the present article.


\section{The Angular Momentum Loss Rate derived by Controlled Models}
\label{sect:controlled}

To prepare to compute the AMLR for our first set of calculations, we generated a grid of models for a range of stellar and disc parameters. We call this set of models “controlled models" because combinations of stellar and disc parameters were prescribed as desired. In these models we used the {\tt SINGLEBE} code \citep{okazaki2007} to create our Be disc models. {\tt SINGLEBE} solves the 1D time-dependent isothermal fluid equations \citep{pringle1981}, under the thin-disc approximation, and assuming subsonic radial velocities for the disc. In our models, the discs form and then dissipate in two years for each cycle, and this is repeated for five cycles (a total length of ten years). Since the average AMLR for the VDD model in \citetalias{ghoreyshi2018} and \citetalias{rimulo2018} are smaller than the predictions of GSE models in \citetalias{granada2013}, we calculated the average AMLR using Equation~\ref{eq:jdot_ave} and the value of 1.0 for $\alpha$, which is upper limit for the VDD model (see Section~\ref{subsect:vdd}). The stellar parameters used are given in Table~\ref{tab:st_param} and are based on the BeAtlas project \citep{faes2015}. In the controlled models we have three free parameters: $W$, DC, and $\Sigma_{0}$ (density at the $R_\mathrm{inj}$). These were chosen because they contribute to the value of the average AMLR through Equations~\ref{eq:sigma_steady}, \ref{eq:jdot_ave}, and \ref{eq:r_eq}. The latter gives $R_\mathrm{eq}$ which in turn is a parameter in Equation~\ref{eq:jdot_steady}. In particular, we adopt values for $W$ of 0.55, 0.75, and 0.95 as a typical lower limit, an average value, and an upper limit, respectively \citep{rivinius2013a}. The DC of 0.25, 0.50, and 0.75 represent a disc with a shorter, equal, and longer formation phases in comparison to the dissipation phase, respectively. Finally, a $\Sigma_{0}$ of 0.80 to 3.2 $\mathrm{g/cm^{2}}$, covers the usual range of disc densities in the Be stars. We note that following the discussion in Section~\ref{sect:models} and according to Figure~\ref{fig:injection_point}, for this Section we adopt an arbitrary value of $1.02 R_\mathrm{eq}$ for $R_\mathrm{inj}$.


\begin{table*}
\begin{center}
\caption{The stellar parameters used in the controlled models of Section~\ref{sect:controlled}, based on the BeAtlas project \citep{faes2015}.}
\begin{tabular}{@{}cccccccccccc}
\hline
\hline
spectral type & \vline & B0 & B1 & B2 & B3 & B4 & B5 & B6 & B7 & B8 & B9 \\
\hline
$M$ ($M_{\odot}$) & \vline & 14.6 & 12.5 & 9.6 & 7.7 & 6.4 & 5.5 & 4.8 & 4.2 & 3.8 & 3.4 \\
$R_\mathrm{pol}$ ($R_{\odot}$) & \vline & 7.50 & 6.82 & 5.80 & 5.11 & 4.62 & 4.26 & 4.02 & 3.72 & 3.55 & 3.37 \\
$T_\mathrm{eff}$ (K) & \vline & 28910 & 26950 & 23630 & 20920 & 18740 & 17060 & 15590 & 14300 & 13330 & 12310 \\
$L$ ($L_{\odot}$ )& \vline & 31183 & 19471 & 8328 & 3971 & 2090 & 1222 & 758 & 460 & 316 & 207 \\

\hline
\end{tabular}
\label{tab:st_param}
\end{center}
\end{table*}

\begin{figure}
    \centering
    \includegraphics[width=\linewidth]{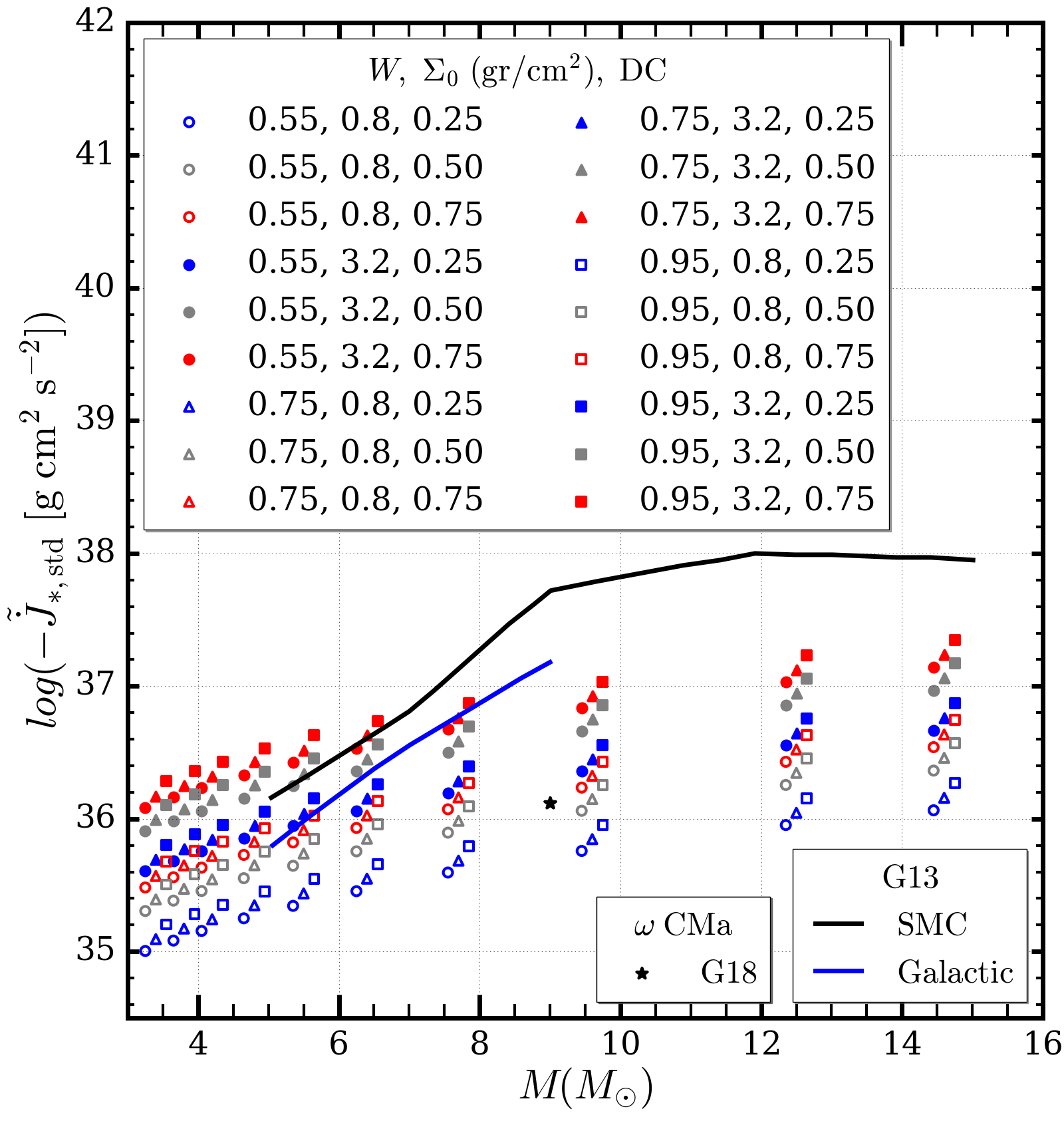}
    \caption{The average AMLR of the model Be stars calculated by the {\tt SINGLEBE} code. Each blue, gray and red symbol is a combination of the stellar parameters and the disc base density as indicated in the legend. The black star shows the average AMLR of the Be star $\omega$ CMa determined by \citetalias{ghoreyshi2018}. The black and blue solid curves give the average AMLR predicted by the GSE model \citepalias{granada2013} for the Be stars in SMC and Milky Way, respectively.}
    \label{fig:am_sinbe}
\end{figure}

Figure~\ref{fig:am_sinbe} shows the average AMLR, $\Tilde{\dot{J}}_{*,\mathrm{std}}$, of the Be stars modelled by the {\tt SINGLEBE} code for different rotational speeds as indicated by the symbols in the Figure. Each color gives an individual value of the DC. The empty and filled symbols represent smaller and larger values of $\Sigma_{0}$, respectively. The solid black and blue curves represent the average AMLR of Be stars predicted by the GSE model for SMC, and Galactic stars, respectively.
There are four main results shown in this figure:
\begin{itemize}
    \item[1)] For B2 type Be stars, combinations of large values of the DC and $\Sigma_{0}$ (even with lower values of $v_\mathrm{rot}$) result in an average AMLR (filled red and gray symbols) close to the predicted values by \citetalias{granada2013}.
    \item[2)] For the mid and late-type Be stars (B3 and later), there are several combinations of $v_\mathrm{rot}$, DC, and $\Sigma_{0}$, according to the average AMLR calculated in this work that are in good agreement with the predicted values by \citetalias{granada2013}.
    \item[3)] The results show that the average AMLR determined for $\omega$ CMa (black star) is closer to the lower limit of the average AMLR calculated using this method. We discuss this later in Section~\ref{sect:discussion}. 
    \item[4)] The overall slope of average AMLR predicted by {\tt SINGLEBE} shows an obvious deviation from that of the GSE model. This may be due to the difference in the free parameters of the two models, e.g, $v_\mathrm{rot}$ of the stars and $\Sigma_{0}$ of the discs. Our models presented in this section include the typical values for these input parameters. We explore a wider range of free parameters in the next section.
\end{itemize}

\section{The Angular Momentum Loss Rate derived by Statistically Selected Calculations}
\label{sect:statistical}

In this section, we present the average AMLR calculated using a set of calculations (SCs) with a wider range of stellar and disc parameters. Similar to the previous section, we use Equation~\ref{eq:jdot_ave} to determine the average AMLR, but this time randomly selected input values given in Table~\ref{tab:param_range} are used to make this calculation. Note, as previously mentioned, the value of $\dot{J}_\mathrm{*, std}$ required in Equation~\ref{eq:jdot_ave} is provided by Equation~\ref{eq:jdot_steady}. This wider range of free parameters gives a large possible stellar and disc parameter combinations. These are more representative sample of average AMLR values. In our SCs, 100 calculations for each sub-spectral type are selected.

The free parameters in our SCs include the stellar mass ($M$), the rotational speed ratio ($W$), duty cycle (DC), stellar polar radius ($R_\mathrm{pol}$), stellar effective temperature ($T_\mathrm{eff}$), and  mass injection rate ($\dot{M}_\mathrm{inj}$). $\dot{M}_\mathrm{inj}$ is the total amount of mass ejected by the star, not just the amount that remains in the disc. There are three more input parameters that are coupled to the free parameters named above. These are $R_\mathrm{eq}$ given by
\begin{equation}
\label{eq:r_eq}
    R_\mathrm{eq} = 0.5(W^{2} + 2)R_\mathrm{pol},
\end{equation}
$R_\mathrm{inj}$ given by
\begin{equation}
\label{eq:r_inj}
    R_\mathrm{inj} = \zeta R_\mathrm{eq},
\end{equation}
where $\zeta$ is a dimensionless parameter with a random value slightly greater than one (given in Table~\ref{tab:param_range}) and, $R_\mathrm{out}$ given by Equation~\ref{eq:r_out}.

\subsection{Statistically Selected Calculations with an Average Rotational Speed}
\label{subsect:1st_model}

In the first SCs (SC1), 100 combinations of input free parameters were randomly selected for each sub-spectral type. However, the upper and lower limits of each free parameter were chosen to match the characteristics of the spectral sub-type given by \cite{harmanec1998} and \cite{granada2014}. For instance, larger values of mass loss rates were used for earlier types in comparison to the later types. The range of each free parameter is given in Table~\ref{tab:param_range}. The values for $W$ are equal for all sub-spectral types. The average AMLR was calculated by Equation~\ref{eq:jdot_ave}. The range for $\dot{M}_\mathrm{inj}$ was calculated using equation 1 of \cite{krticka2014} based on the critical values of mass and temperature for each sub-spectral type.

Also, observations suggest that the discs of late-type Be stars are more stable than the earlier types \citep{labadie2018}. Therefore, we considered larger values of the DC for late-type Be stars. Thus, while the DC for the early-type stars were limited between 0.3 and 0.6, we considered a narrower and larger range of DC for the late-type Be stars (B7 and later) in the SCs, i.e., between 0.9 and 1.0. A value of DC near 1.0 means that the star is constantly feeding the disc.


\begin{table*}
\begin{center}
\caption{The range of free parameters for the star and disc used in the statistically selected SCs.}
\begin{tabular}{@{}cccccccccccc}
\hline
\hline
spectral type & \vline & B0 & B1 & B2 & B3 & B4 & B5 & B6 & B7 & B8 & B9 \\
\hline
$M$ ($M_{\odot}$) & \vline & 13.5-17.5 & 11.0-13.5 & 8.7-11.0 & 7.0-8.7 & 6.0-7.0 & 5.2-6.0 & 4.5-5.2 & 4.0-4.5 & 3.6-4.0 & 3.2-3.6 \\
$R_\mathrm{pol}$ ($R_{\odot}$) & \vline & 7.15-8.00 & 6.30-7.15 & 5.46-6.30 & 6.0-7.0 & 4.44-4.87 & 4.14-4.44 & 3.87-4.14 & 3.64-3.87 & 3.46-3.64 & 3.28-3.46 \\
$T_\mathrm{eff}$ (kK) & \vline & 25-32 & 21-25 & 18.5-21.0 & 16.5-18.5 & 14.500-16.5 & 13.5-14.5 & 12.5-13.5 & 11.5-12.5 & 10.5-11.5 & 9.5-10.5 \\
$\dot{M}_\mathrm{inj}$ ($10^{-7} M_{\odot}/\mathrm{yr}$ )& \vline & 0.035-35 & 0.031-20 & 0.024-10 & 0.02-6 & 0.017-2.7 & 0.015-1.2 & 0.013-0.52 & 0.002-0.2 & 0.001-0.1 & 0.0005-0.05 \\
$\zeta$ & \vline & 1.01-1.1 & 1.01-1.1 & 1.01-1.1 & 1.01-1.1 & 1.01-1.1 & 1.01-1.1 & 1.01-1.1 & 1.01-1.1 & 1.01-1.1 & 1.01-1.1 \\
$W$ in SC1 & \vline & 0.55-0.95 & 0.55-0.95 & 0.55-0.95 & 0.55-0.95 & 0.55-0.95 & 0.55-0.95 & 0.55-0.95 & 0.55-0.95 & 0.55-0.95 & 0.55-0.95 \\
$W$ in SC2 & \vline & 0.90-0.95 & 0.90-0.95 & 0.90-0.95 & 0.90-0.95 & 0.90-0.95 & 0.90-0.95 & 0.90-0.95 & 0.90-0.95 & 0.90-0.95 & 0.90-0.95 \\
$\mathrm{DC}$ in SC1 & \vline & 0.30-0.60 & 0.30-0.60 & 0.30-0.60 & 0.30-0.60 & 0.30-0.60 & 0.30-0.60 & 0.30-0.60 & 0.90-1.00 & 0.90-1.00 & 0.90-1.00 \\
$\mathrm{DC}$ in SC2 & \vline & 0.90-1.00 & 0.90-1.00 & 0.90-1.00 & 0.90-1.00 & 0.90-1.00 & 0.90-1.00 & 0.90-1.00 & 0.90-1.00 & 0.90-1.00 & 0.90-1.00 \\

\hline
\end{tabular}
\label{tab:param_range}
\end{center}
\end{table*}


According to \citetalias{rimulo2018} and \citetalias{ghoreyshi2018}, the values of the $\alpha$ parameter are different during disc build up and dissipation, with larger values for disc formation (close to 1.0 for disc build up and about 0.20 for disc dissipation). In this work, the value of $\alpha$ was set to 1.0. However, we should note for these 1D SCs that $\alpha$ controls only the time i.e., a disc with a higher $\alpha$ evolves faster than a disc with a lower $\alpha$. Our SCs in this Section are calculated for a wide range of the DC. The average AMLR based on other values of $\alpha$ can be achieved by varying the DC. For example, if we consider a smaller value of $\alpha$ for a disc dissipation phase, this would result in a longer disc dissipation phase, and consequently, a smaller DC, which is considered in our grid.

\begin{figure}
    \centering
    \includegraphics[width=\linewidth]{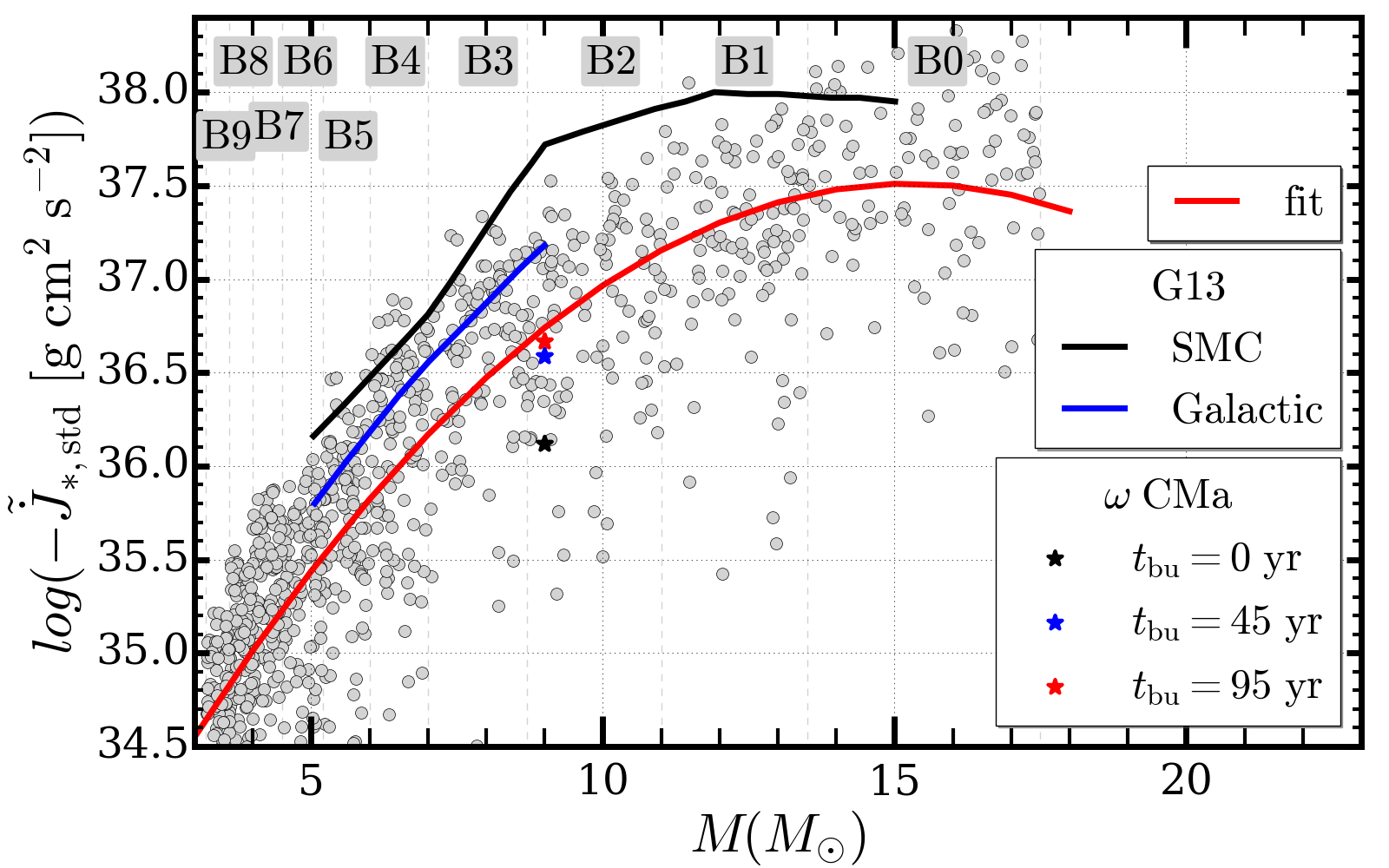}
    \caption{The average AMLR (gray circles) versus stellar mass for SC1, \citetalias{granada2013}, and \citetalias{ghoreyshi2018}. The red solid line shows our fit to the average AMLR values from SC1. The black and blue solid curves show the AMLR of Be stars predicted by the GSE model \citepalias{granada2013}. The black star shows the average AMLR of the Be star, $\omega$ CMa \citepalias{ghoreyshi2018} without considering the 45-year disc build up. The blue and red stars show the average AMLR of the star $\omega$ CMa in this work, considering a 45-year and a 95-year disc build up, respectively.}
    \label{fig:am1}
\end{figure}

The results of SC1 are provided  and compared with the predictions of \citetalias{granada2013} in Figure~\ref{fig:am1}. The gray circles show the average AMLR for each star. The black and blue curves show the average AMLR of the stars in SMC and Milky Way predicted by \citetalias{granada2013}, respectively. The red curve is our model fit. The average AMLR calculated for $\omega$ CMa by \citetalias{ghoreyshi2018} is shown by a black star. Similar to the controlled SCs presented in Section~\ref{sect:controlled}, the red curve confirms that the discrepancy between the values predicted by \citetalias{granada2013} and calculated by this work is less than what was reported by \citetalias{ghoreyshi2018} for $\omega$ CMa.

\subsection{Statistically Selected Calculations with a Rotational Speed Close to Critical}
\label{subsect:2nd_model}

\begin{figure}
    \centering
    \includegraphics[width=\linewidth]{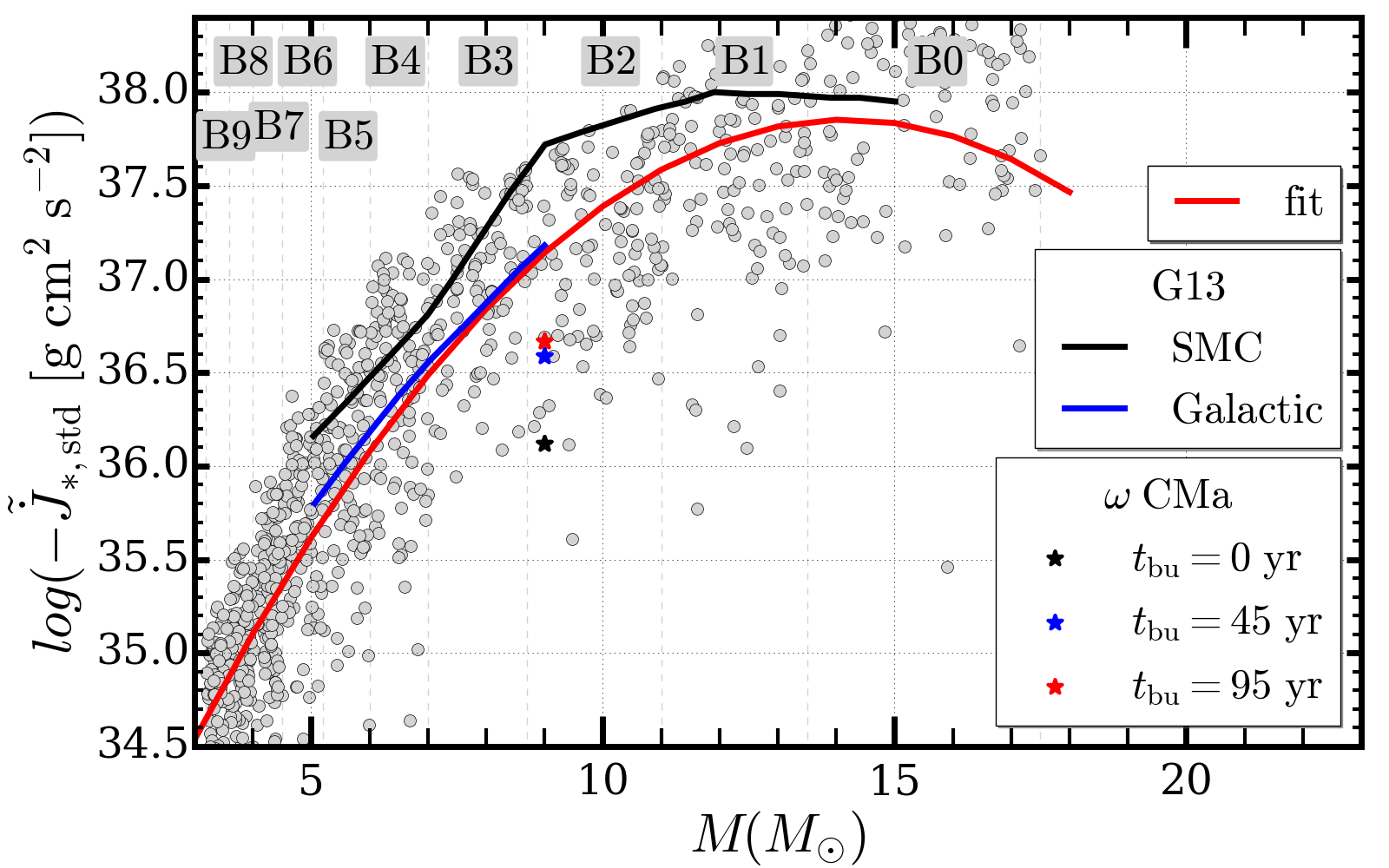}
    \caption{Same as Figure\,\ref{fig:am1} for SC2.}
    \label{fig:am2}
\end{figure}

As mentioned in Section~\ref{sect:intro}, according to the GSE model mechanical mass loss occurs when the star reaches $W=0.88$, which leads to the formation of a Be star disc. This value of $W$ is greater than the average rotational speed of $W=0.75$ reported for Be stars  \citep[see figure 9 in][and the references in the caption of the figure]{rivinius2013a}. Therefore, in order to compare VDD models with the same conditions as those proposed for the GSE calculations, in our second set of SCs (SC2) we restricted the $W$ parameter to the values close to the critical velocity. Also, DC was set to be close to 1.0 for all SCs to represent the stars that feed the disc, continuously. The limits for all other free parameters are the same as SC1. The range of each free parameter is given in Table~\ref{tab:param_range}. The results of SC2 are shown and compared by the predictions of \citetalias{granada2013} in Figure~\ref{fig:am2}. For our assumptions in SC2 there are no discrepancies between the results of the VDD and \citetalias{granada2013}. This confirms that the source of difference between the AMLR values predicted by the VDD and GSE models is the fact that near-critical stellar rotation was assumed by the GSE model.

\section{Discussion}
\label{sect:discussion}

\begin{figure}
    \centering
    \includegraphics[width=\linewidth]{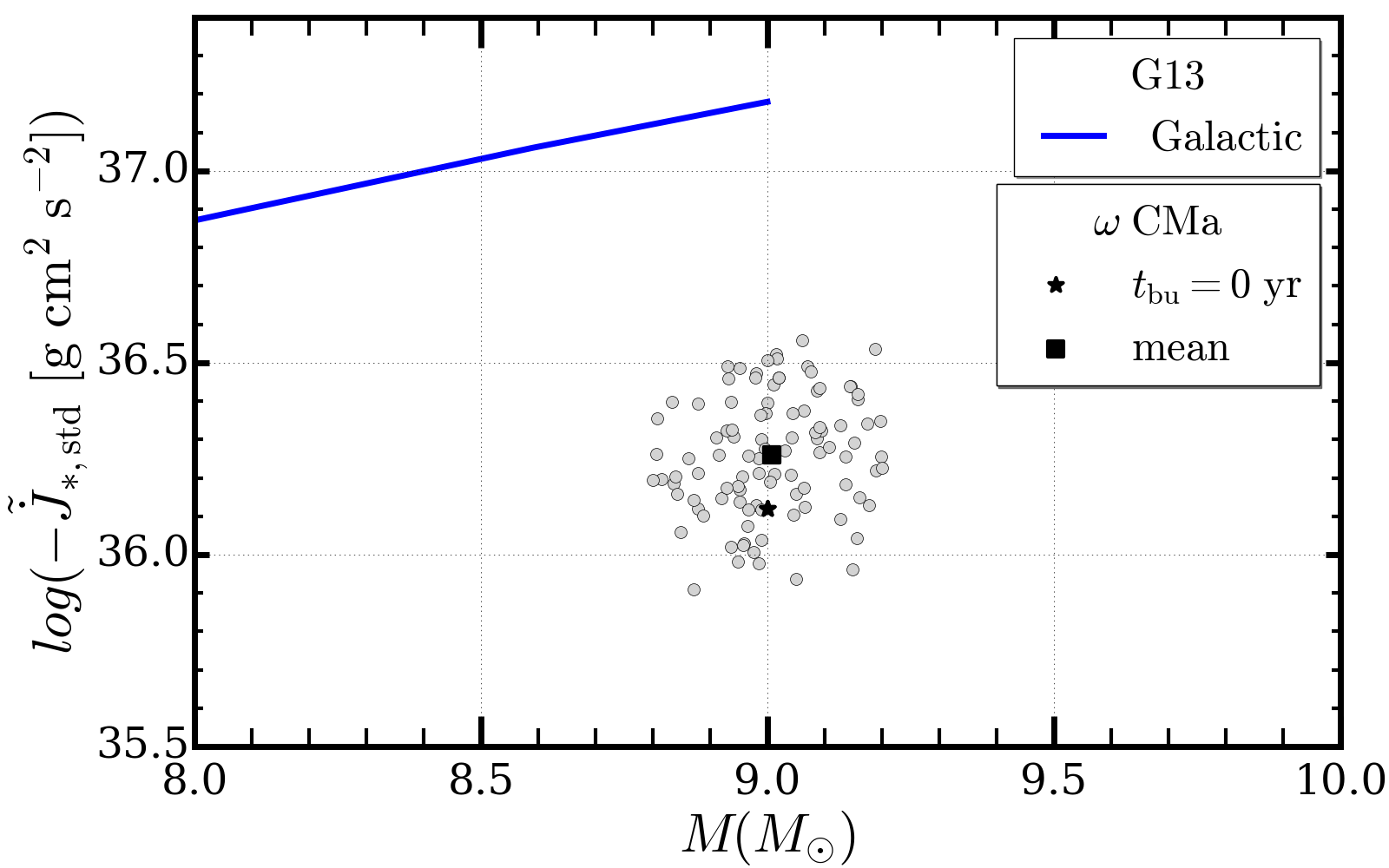}
    \caption{The average AMLR value for the star, $\omega$ CMa, as calculated by \citetalias{ghoreyshi2018} (black star) compared with the average AMLR values calculated for a sample Be stars with stellar parameters close to $\omega$ CMa (gray circles) and the values predicted by \citetalias{granada2013} (blue line). The mean value of the AMLR for the gray circles is shown by the black square.}
    \label{fig:omecma}
\end{figure}

Although SC2 explains differences between the average AMLR predicted by \citetalias{granada2013} and the average AMLR determined by \citetalias{rimulo2018} and \citetalias{ghoreyshi2018}, it is still remarkable that the average AMLR for the B2e star, $\omega$ CMa determined by \citetalias{ghoreyshi2018} is close to the lower limit for Be stars of similar mass. To investigate whether the average AMLR for $\omega$ CMa could have been underestimated, we calculated the average AMLR (using Equations~\ref{eq:jdot_steady}, and \ref{eq:jdot_ave}) for a sample of 100 B2-type Be stars. The stellar parameters for B2 type stars in this sample were chosen over a narrower range of mass (between 8.8 and 9.2$M_\odot$) and a polar radius (between 5.9 and 6.1$R_\odot$), to restrict the SCs to stars most similar to $\omega$ CMa with $M=9.0M_\odot$ and $R_\mathrm{pol}=6.0R_\odot$. Then, we calculated the mean value of average AMLR for this sample as shown in Figure~\ref{fig:omecma}. The black star shows the average AMLR for $\omega$ CMa in \citetalias{ghoreyshi2018}. For this restricted sample, the mean value of the average AMLR is slightly larger than the value reported in \citetalias{ghoreyshi2018}. It may be due to the fact that the $V$-band photometric data of $\omega$ CMa that were used in \citetalias{ghoreyshi2018} were sparse for most of the disc formation phases and these phases are a significant part of the disc evolution for calculating the average AMLR. Therefore, the value of average AMLR could be underestimated.


\citetalias{ghoreyshi2018} included the mass reservoir (mass accumulated in the outer parts of the disc) effect \citepalias{rimulo2018, ghoreyshi2018} in the modelling process, but they did not include it in the average AMLR calculations. They found that ignoring the mass reservoir effect in the work by \cite{carciofi2012} caused an overestimation of $\alpha$. Also, \citetalias{ghoreyshi2018} showed that the mass reservoir effect is an essential parameter for modelling Be stars since it can feed the inner disc during the disc dissipation. To include this effect, they assumed a long disc formation phase ($t_\mathrm{bu}$) of 45 years followed by a 5-year dissipation (typical for $\omega$ CMa) before the first cycle of brightness variations in their modelling. In the 34-year period that was modelled for $\omega$ CMa, the mass reservoir was always present and continuously fed the inner regions of the disc. Having this large mass reservoir is reasonable if we consider that $\omega$ CMa has an age in the order of million years during which the $t_\mathrm{bu}$ has had a typically slightly shorter ($\approx$ 80\%) length, but significantly larger values of $\alpha$ parameter ($\approx$ 3 to 5 times) in comparison to the disc dissipation phase \citepalias[see figure 11 in][]{ghoreyshi2018}. Therefore, this long $t_\mathrm{bu}$ resulted in accumulation of a significant amount of mass at large disc radii which remained during periods when mass ejection was turned off, and it affected the average AMLR.

\begin{figure}
    \centering
    \includegraphics[width=\linewidth]{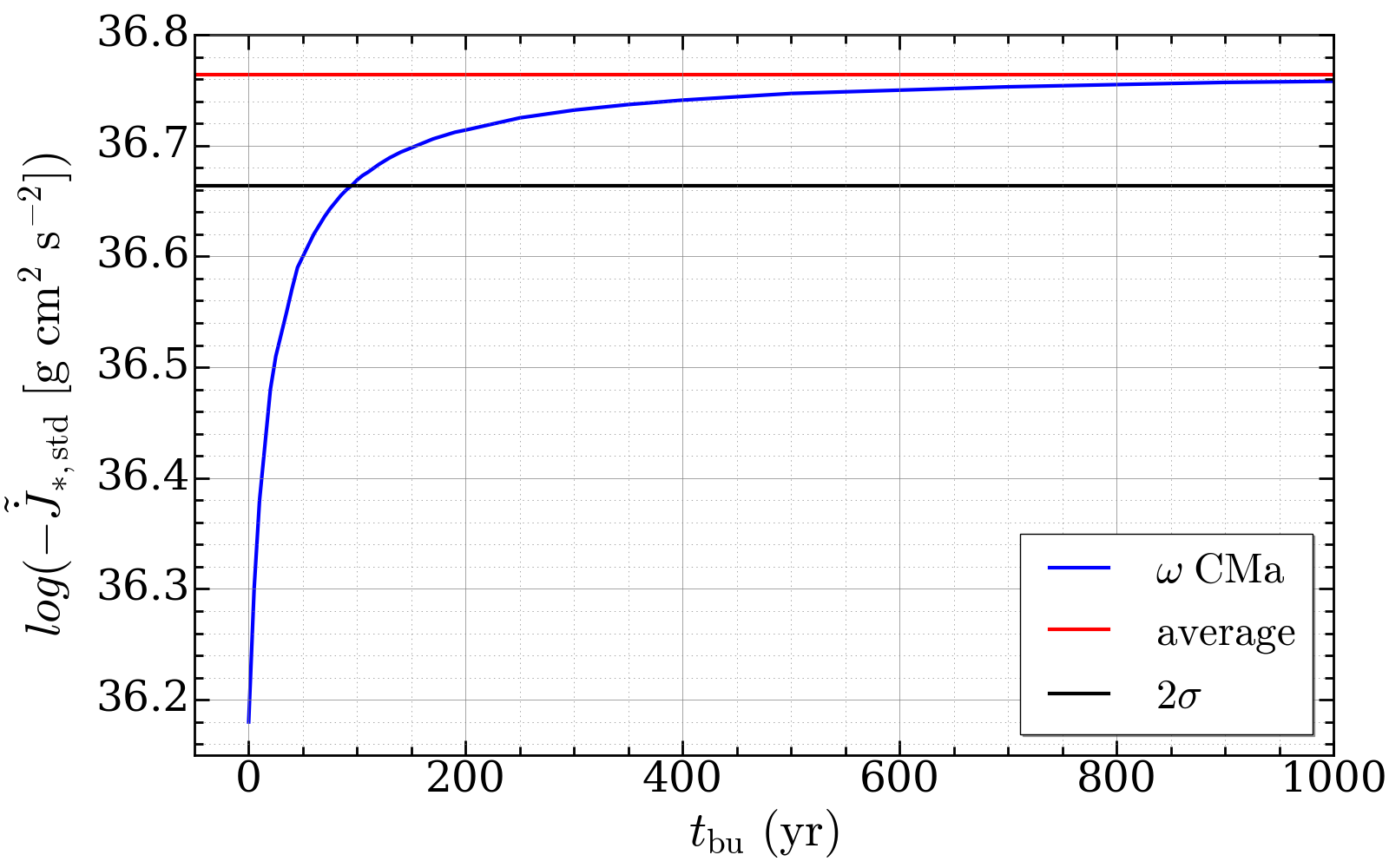}
    \caption{The average AMLR for $\omega$ CMa versus the length of disc build up time (blue curve). The red line shows the average AMLR for typical B2-type stars from our SC1. The black line shows a $2\sigma$ deviation from the upper limit of average AMLR for $\omega$ CMa.}
    \label{fig:j_vs_dc}
\end{figure}

To calculate the average AMLR for $\omega$ CMa, \citetalias{ghoreyshi2018} first used Equation~\ref{eq:jdot_steady} to obtain the AMLR for each epoch during the 34-year life cycle of the star. Then, they averaged the AMLR over this time span by using Equation~\ref{eq:jdot_ave_integral}, which resulted in $1.2\times 10^{36}\,\mathrm{g\,cm^2\,s^{-2}}$, as shown by the black star in Figures~\ref{fig:am_sinbe} to \ref{fig:omecma}. \citetalias{ghoreyshi2018} included the 45-year $t_\mathrm{bu}$ in all their modelling but did not include this particular detail in the calculation of the average AMLR. We have found that in order to calculate the average AMLR, $t_\mathrm{bu}$ must be included. Therefore, we recalculated the average AMLR for $\omega$ CMa, using the 45-year disc formation and 5-year disc dissipation and determined a new value of $3.9\times 10^{36}\,\mathrm{g\,cm^2\,s^{-2}}$ (the blue star in Figures~\ref{fig:am1} and \ref{fig:am2}). This value is still slightly smaller than the average AMLR for a typical B2-type star with the stellar parameters given in Table~\ref{tab:st_param}. Then, we investigated what the required $t_\mathrm{bu}$ of $\omega$ CMa would be if we assumed its average AMLR approaches the average value obtained for a typical B2-type star. It is important to note that due to differences between the stellar parameters of  $\omega$ CMa and a typical B2 star, we do not expect to reach the average AMLR for a typical B2-type star, but rather explore what the $t_\mathrm{bu}$ would be if we considered an average AMLR deviating $2\sigma$ from such a value (the black line in Figure~\ref{fig:j_vs_dc}). We obtain a $t_\mathrm{bu} = 95$ years and an average AMLR for $\omega$ CMa of $4.7\times 10^{36}\,\mathrm{g\,cm^2\,s^{-2}}$, as indicated by the red star in Figure~\ref{fig:am1}.


\begin{figure}
    \centering
    \includegraphics[width=\linewidth]{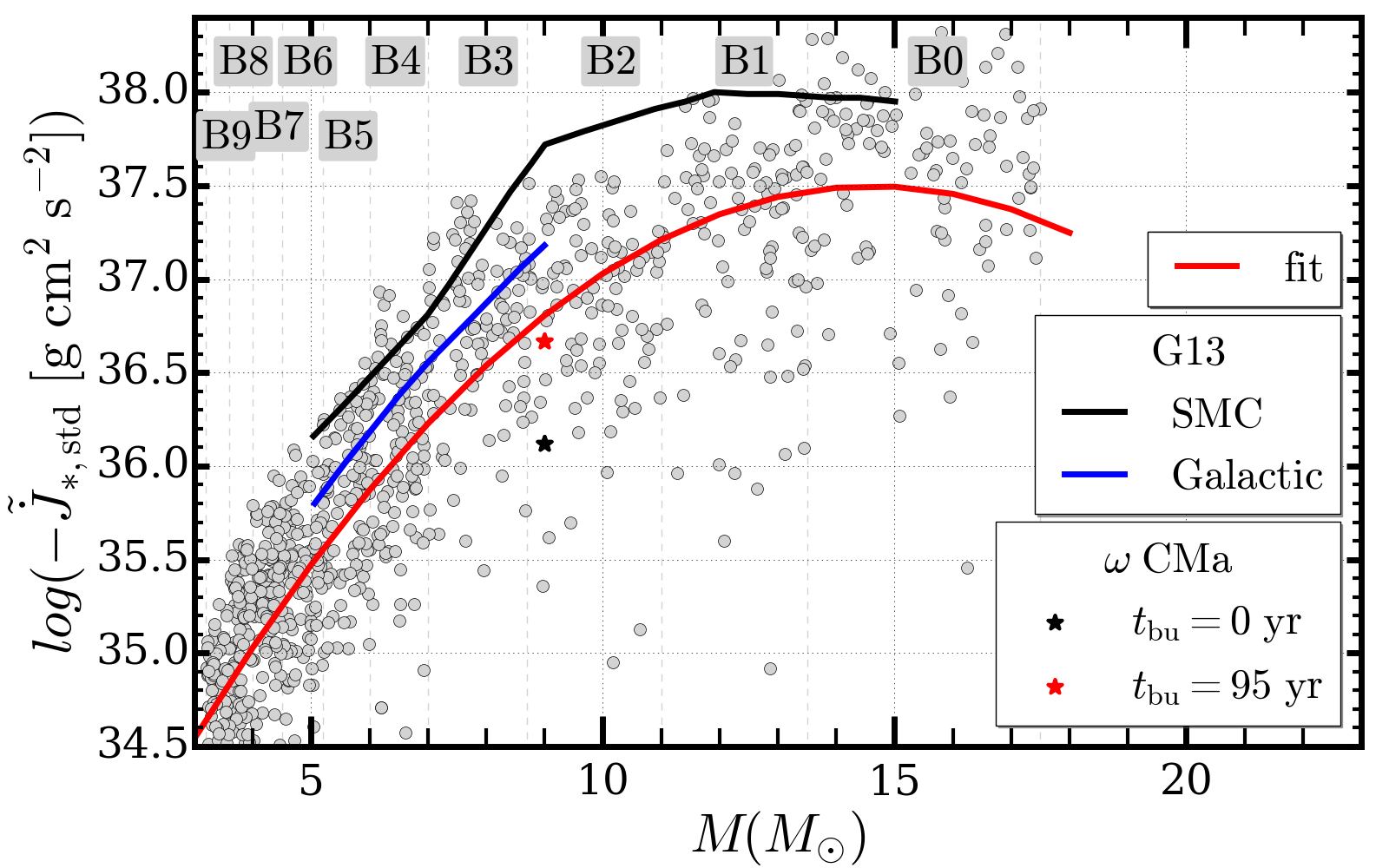}
    \caption{Same as Figure\,\ref{fig:am1} for a smaller range of  $R_\mathrm{out}$, to represent truncated discs in binary systems.}
    \label{fig:truncation}
\end{figure}

Finally, it is worth discussing the effect of truncated discs, due to interactions of a companion, on the average AMLR. Studying more than 400 B-type stars in the Large Magellanic Cloud, \citet{dunstall2015} found that about half of these were likely a member of binary, triplet, etc. If the semi-major axis of the binary system is large, then, the outer radius of the disc could be large, too. As we discussed in Section~\ref{subsect:vdd}, in such a case, the average AMLR depends slightly on the outer radius of the disc. But, what if the outer radius is not very large (i.e., truncated by a close companion)? Is its influence on the average AMLR, still insignificant? Although the sharp truncation of the Be discs by close binaries is not commonly observed \citep{klement2019}, we decided to investigate to see what this analysis would reveal about the AMLR. To address this question, we repeated the calculations presented in Section~\ref{subsect:1st_model}, for smaller outer radii between 15 and 75 $R_\mathrm{eq}$. The results are shown in Figure~\ref{fig:truncation}, but the difference compared with Figure~\ref{fig:am1} is small. We can barely see that the red solid line (representing the fit for the statistical SCs predicted by the VDD paradigm) in Figure~\ref{fig:truncation} became closer to the blue solid line (representing the predicted values of average AMLR by the GSE model for the stars in Milky Way). Thus, if the presence of small truncated discs in the Be stars is relevant, this decreases the discrepancies between the VDD and GSE models, insignificantly.



\begin{figure*}
    \begin{minipage}{0.5\linewidth}
        \centering
        \subfloat[B2-type]{\includegraphics[width=1.0\linewidth]{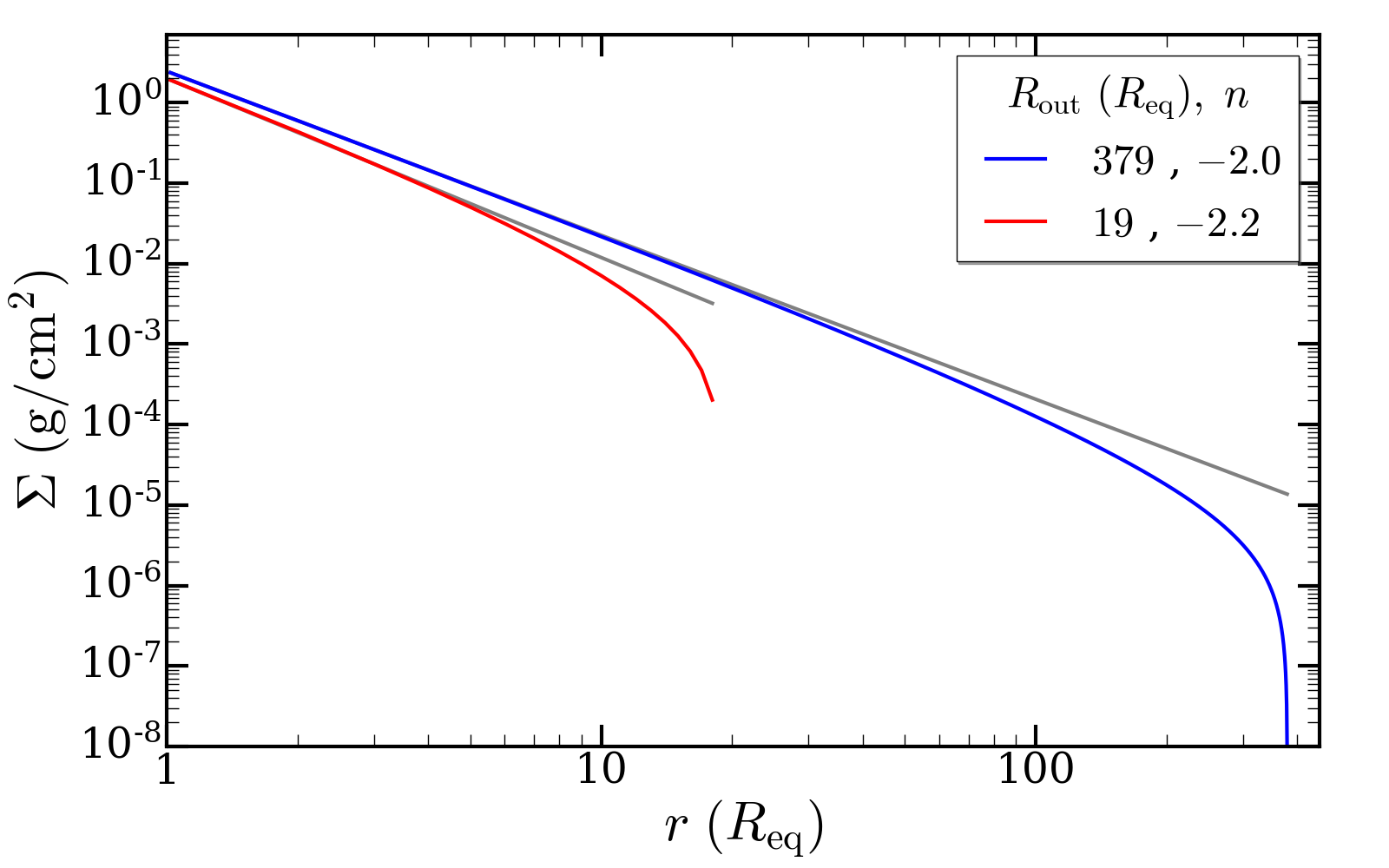}}
    \end{minipage}%
    \begin{minipage}{0.5\linewidth}
        \centering
        \subfloat[B7-type]{\includegraphics[width=1.0\linewidth]{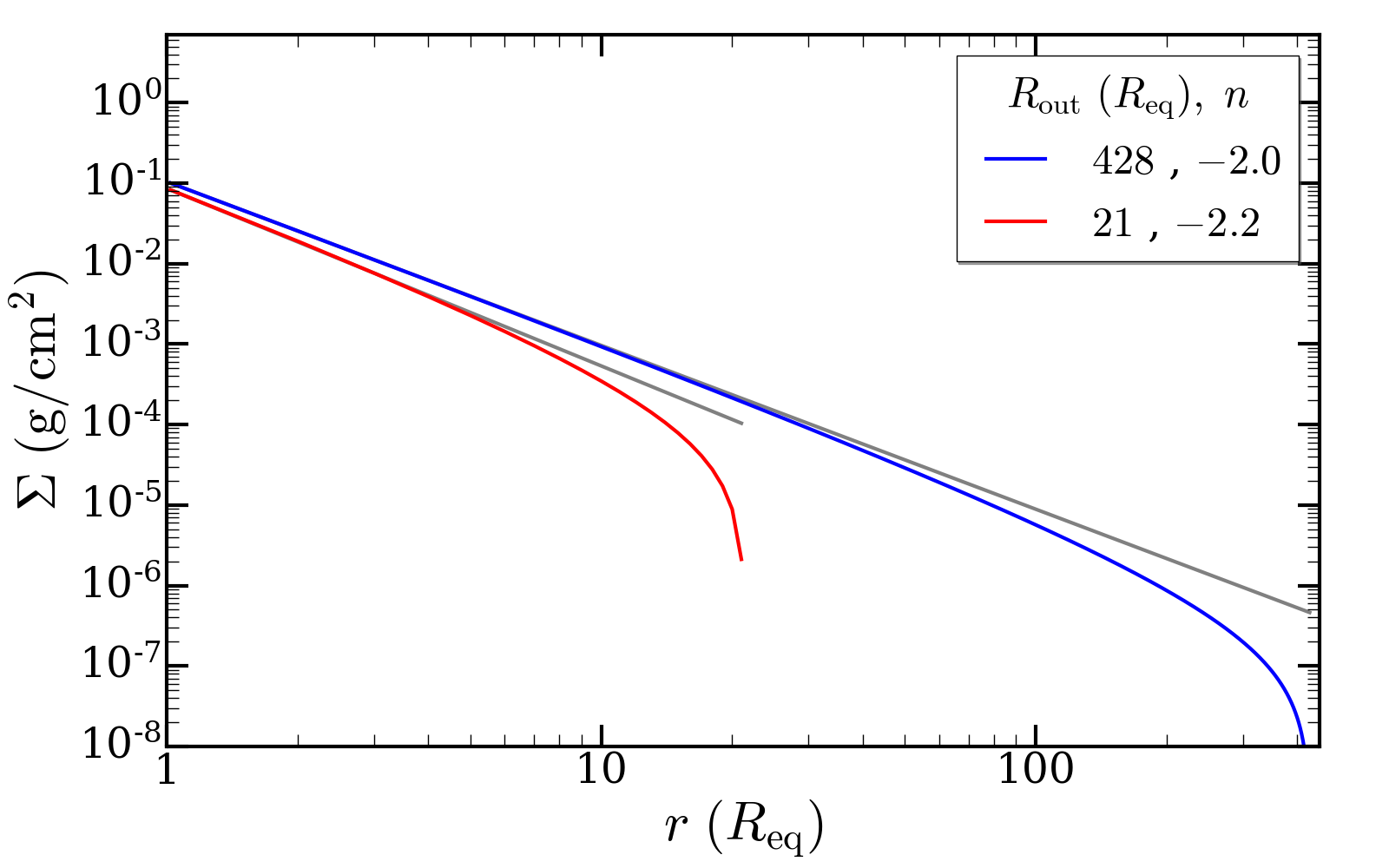}}
    \end{minipage}
    \caption{Effect of disc truncation on the disc surface density profile slope if the AMLR remains constant for an early-type Be star (left panel) and a late-type Be star (right panel). The blue and red line show the disc surface density as a function of radius for a large and truncated disc, respectively. The gray lines are the tangents of each curve, which represent the slope of the disc surface density profile.}
    \label{fig:surface_density}
\end{figure*}

Although the disc truncation does not seem to change the average AMLR of the Be stars, it does change the disc surface density slope. Figure~\ref{fig:surface_density} shows how the slope of disc surface density profile changes by disc truncation. The left and right panels of the figure show the disc surface density versus radius for an early-type (B2) and a late-type (B7) Be star with the main stellar parameters given in Table~\ref{tab:st_param}, $W=0.75$, $\zeta=1.02$, and $\mathrm{DC}=0.50$ for both stars, and $\dot{M}_\mathrm{inj}=10.0\times10^{-7}$ and $0.20\times10^{-7} M_{\odot}/\mathrm{yr}$, for B2 and B7 type stars, respectively. The blue and red lines correspond to a large and truncated disc, respectively. The gray lines are the tangents of each curve, which represent the slope of disc surface density profile, $n$, for a power law fall-off. The AMLR was kept fixed for each sub-spectral type and Equation~\ref{eq:sigma_steady} was used for calculating the disc surface density. The outer radius of the disc, $R_\mathrm{out}$, was calculated using Equation~\ref{eq:r_out} and the size of truncated disc was assumed to be 5\% of $R_\mathrm{out}$. We see that the magnitude of the power law slope, $n$, increases from 2.0 (representing a steady-state disc) for a large disc to 2.2 for a truncated disc, regardless of the sub-spectral type of the Be star. 

Finally, it is worth mentioning that an important fraction of Be stars do not show significant variability \citep[][and references therein]{rivinius2013a}. For many of these stable stars, in particular those rotating close to the critical limit, the AMLRs predicted by \citetalias{granada2013} are likely correct.


\section{Conclusions}
\label{sect:conclusions}

We use the VDD model to calculate the average AMLR for a large sample of Be stars. Our work includes Be stars with a wide range stellar and disc parameters. By comparing our results with the predictions of the GSE models:

\begin{itemize}
    \item The main hypothesis for triggering mechanical mass loss in the GSE models is rotation of the Be stars close to the critical speed. We show that the average AMLR predictions of the VDD model and \citetalias{granada2013} are in agreement when Be stars rotate close to their critical speeds. We note the late-type Be stars are already in a good agreement even with more moderate rotation speeds (Figure~\ref{fig:am_sinbe}).
    \item The history of the star and disc play an important role in the reservoir of mass accumulated at large radial distances, which in turn, affects the average AMLR.
    \item We investigate the role of disc truncation by a binary companion in a Be star on the average AMLR and find this has a minor effect.
    \item For the first time, we determine a range of average AMLR between $3.0\times 10^{34}$ to $5.0\times 10^{35}\,\mathrm{g\,cm^2\,s^{-2}}$ for the late-type Be stars. This range is much larger for the early-type Be stars, which is mainly due to the greater range in their stellar parameters (see Table~\ref{tab:param_range}). We note that the late-type Be stars are not well studied in the literature and further investigations on this group of stars are likely to provide a fresh insight of this subject.
\end{itemize}

\section*{Acknowledgements}
The authors would like to thank the anonymous referee for their very thorough and detailed comments and suggestions that improved the paper. M.R.G. acknowledges the grant awarded by the Western University Postdoctoral Fellowship Program (WPFP). C.E.J. acknowledges the Natural Sciences and Engineering Research Council of Canada for the financial support. A.G. acknowledges the financial support received from the Universidad Nacional de Rio Negro, Argentina (PI2020-UNRN 40B-890).

\section*{Data availability}
No new data were generated or analysed in support of this research.


\bibliographystyle{mnras}
\bibliography{references}


\bsp	
\label{lastpage}
\end{document}